\documentstyle[12pt]{article}

\newcommand{\be}{\begin{equation}}
\newcommand{\ee}{\end{equation}}
\newcommand{\bd}{\begin{displaymath}}
\newcommand{\ed}{\end{displaymath}}
\newcommand{\baa}{\begin{array}{lll}}
\newcommand{\eaa}{\end{array}}
\newcommand{\ba}{\begin{eqnarray}}
\newcommand{\ea}{\end{eqnarray}}

\begin{document}

\begin{center}
{\LARGE\bf Transverse spin effects and higher twist in the singlet channel}

\vspace{1cm}
{J. Soffer$^a$,  O.V. Teryaev$^b$}

\vspace*{1cm}
{\it $^a$Centre de Physique Th\'eorique - CNRS - Luminy,\\
Case 907 F-13288 Marseille Cedex 9 - France}\\

\vspace*{3mm}
{\it $^b$Bogoliubov Laboratory of Theoretical Physics, \\
Joint Institute for Nuclear Research, Dubna, 141980, Russia}\\

\vspace*{2cm}

\end{center}

\begin{abstract}
The effects, originating from the longitudinal gluon
polarization, are considered.
We derive the analogue of the Wandzura-Wilczek relation for the light-cone
distributions of polarized gluons in a transversely polarized nucleon.
The short distance cross-section is entirely due to the intrinsic
transverse momenta of gluon in the nucleon, in complete analogy
to the quark case. Numerical estimate for the double spin transverse
asymmetries are presented.
The distribution of longitudinal gluons in a nucleon,
introduced earlier by Gorsky and Ioffe, is
estimated from below by making use of the positivity of density matrix.
The similar upper bound for the nonperturbative
polarized charm is discussed.

\end{abstract}

\section{Introduction}

\vspace{1mm}
\noindent

   It is now well known that the spin properties of gluons and quarks are
fairly different. In particular,
there is no analogue of the twist-two
transversity distributions for massless gluons and  their contribution to the
transverse asymmetry starts at the twist-three level.
Also, longitudinal massless gluons do not exist.
However, due to the confinement property, gluons should acquire
an average mass and/or a transverse momentum of the order of the inverse
of the
hadron size.
As a result, one can have a nonzero longitudinal gluon distribution.
Generally speaking, it is suppressed by the gluon mass squared and
contributes
at the twist-four  level.
However, in the case of Deep Inelastic Scattering (DIS)
this is cancelled by the pole in the box diagram.
This effect was studied in details for longitudinal gluons
by Gorski and Ioffe \cite{GI}.
It was shown to be related to the conformal anomaly,
and one may wonder, if it could be observed in other processes.

The gluon mass and/or its intrinsic transverse momentum
should result also in a nonzero transverse gluon distribution.
It was recently studied in \cite{ST97}
where the twist-two approximation was derived.
The resulting double transverse spin asymmetries $A_{TT}$, for low mass dijet
or low $p_T$ direct photon production,
at RHIC are rather small ($\leq 1\%$) due to a kinematic suppression factor.
It seems also promising to study the double
asymmetries
in  open charm or heavy quarkonia leptoproduction
by a longitudinally polarized lepton beam off a transversely polarized
target. The suppression factor that enters in this case is to the first power
of $M/\sqrt{\hat s}$, contrary to the second power for some double
transverse spin asymmetries. The situation seems especially favorable
for the  case of  diffractive
charmonium production,investigated recently \cite{RY},
when the partonic c.m. energy $\sqrt{\hat s}$
is of the order of the charm quark mass.
This would make possible the measurement of the transverse gluon distribution.

The present paper is devoted to the discussion of the manifestaion
of longitudinal gluons
in the polarized and unpolarized nucleons and the relations between them.

\section{Gluonic contribution to the transverse
polarization of nucleon}

\vspace{1mm}
\noindent

  The absence of the twist-two transversity distribution
for gluons, mentioned above,is leading
generally speaking, to the relative suppression
of gluon transverse asymmetries with respect to the quark ones,
which was used recently to formulate the selection rule \cite{JaSa}
in QCD, which is of direct relevance for the physics programme
of the future polarized $pp$ collider at BNL-RHIC.

However, the detailed analysis of the quark contribution to the
double transverse spin asymmetries $A_{TT}$
using the Monte-Carlo simulation \cite{MS}, resul\-ted in rather small numbers
of the order of $1\%$, and therefore it seems natural to
question the role of gluon corrections.

The transverse polarization effects are arising from two basic
sources: the leading twist transversity distribution, resulting
in the correlation of transverse polarizations, and the twist-three
parton correlations, suppressed by the hadron mass. While the first are
absent for gluons, the gluon cor\-re\-la\-tions are, generally speaking,
rather complicated \cite{Ji}.
At the same time, the experimental data on the $g_2$ structure
function \cite{g2}
do not deviate strongly from the twist-two approximation,
suggested by Wandzura and Wilczek (WW) \cite{WW}, whose physical meaning
is just the dominance of the effect of transverse motion of the quark
over that of the gluon field \cite{kt}.
This is the reason, why we are suggesting here
a generalization of the WW approximation  to
the case of gluons.

To do this, let us start from the light-cone density matrix of gluon,
namely:
\begin{eqnarray}
\label{dm}
\int{{d\lambda \over2\pi}}
e^{i\lambda x}
\langle p,s|A^\rho(0) A^\sigma (\lambda n) |p,s \rangle \sim
{1 \over 2} (e_{1T}^\rho e_{1T}^\sigma+e_{2T}^\rho e_{2T}^\sigma)G(x)
+\nonumber\\
{i \over 2} (e_{1T}^\rho e_{2T}^\sigma-e_{2T}^\rho e_{1T}^\sigma)\Delta G (x)
+
{i \over 2} (e_{1T}^\rho e_L^\sigma-e_L^\rho e_{1T}^\sigma)
\Delta G_T (x) + G_L(x) e_L^\rho e_L^\sigma~,
\end{eqnarray}
where $n$ is the gauge-fixing light-cone vector such that $np=1$,
and we define two transverse polarization vectors,
$e_{1T}$ and $e_{2T}$ . One of them, namely $e_{2T}$ for definiteness,
is chosen to be parallel to the direction of the transverse component of the
polarization, so that's why only the vector $e_{1T}$ enters in the
contribution of $\Delta G_T (x)$. Also, we introduce the longitudinal
polarization vector
$e_L$.
We denote by
$s_{\mu }$ the covariant
polarization vector  of the proton of momentum
$p$ and mass $M$ and we have $s^2=-1, sp=0$.
Here $G(x)$ and $\Delta G(x)$ are the familiar unpolarized gluon distribution
and gluon helicity distribution, respectively.
The transverse gluon distribution $\Delta G_T(x)$
is the most natural way to measure its transverse polarization,
analogous to the quark structure function $g_T=g_1+g_2$,
since in the quark case we have:

\begin{eqnarray}
\label{quark}
{1\over{4M}}\int{{d\lambda \over2\pi}}e^{i\lambda x}
\langle p,s|\bar \psi (0)\gamma^{\mu}
\gamma ^5\psi (\lambda n)|p,s \rangle=g_1(x)(sn)p^{\mu}+g_T (x) s_T^{\mu}~.
\end{eqnarray}

The quantity
$g_T$ was shown
to be the good variable to study the generalized Gerasimov-Drell-Hearn
sum rule, and the $x$ dependence of the
anomalous gluon contribution \cite{ST95}.
The latter result was recently confirmed \cite{rav}.

The light-cone distributions $\Delta G$ and $\Delta G_T$ can be
easily obtained
by the projection of gluon density matrix, so we have,

\begin{eqnarray}
\label{distr}
\Delta G(x)={ix\over{4M(sn)}} \int{{d\lambda \over2\pi}}
e^{i\lambda x}
\langle p,s|A_\rho(0) A_\sigma (\lambda n) |p,s \rangle
\epsilon ^{\rho \sigma p n}, \nonumber \\
\Delta G_T(x)={ix\over{4M(s^2)}} \int{{d\lambda \over2\pi}}
e^{i\lambda x}
\langle p,s|A_\rho(0) A_\sigma (\lambda n) |p,s \rangle
\epsilon ^{\rho \sigma p s}.
\end{eqnarray}

Now by making use of the axial gauge $An=0$, one may express their moments
\footnote {The first moment requires to take into account the non-local
operator \cite{ET88,BB}.At the same time, the
non-local operators found in the renormalization of
the gluon contribution to $g_2$ \cite{BET94}
should be equal to zero, when one uses the gauge invariance and
equations of motion.}
in terms of gluon field
strength $G_{\mu \nu}$, according to

\begin{eqnarray}
\label{distrG}
\int_0^1 dx x^k \Delta G(x)={1\over{4M(sn)}}
\langle p,s|G_{\rho n}(0) (i \partial n)^{k-1} G_{\sigma n}(0) |p,s \rangle
\epsilon ^{\rho \sigma p n}, \nonumber \\
\int_0^1 dx x^k \Delta G_T(x)={1\over{4M(s^2)}}
\langle p,s|G_{\rho n}(0) (i \partial n)^{k-1} G_{\sigma n}(0) |p,s \rangle
\epsilon ^{\rho \sigma p s}.
\end{eqnarray}

We denote here $G^{\mu n}=G^{\mu \nu} n_{\nu},
\partial n=n_{\mu}\partial^{\mu}$, and we recall that in configuration space
$x^k=(i \partial n)^k$.
The kinematical identities, implied by the vanishing of the
totally antisymmetric tensor of rank 5 in four-dimensional space,

\begin{eqnarray}
\label{kin}
n^{\mu}\epsilon_{\rho \sigma p n}-
n^{\sigma}\epsilon_{\rho \mu p n}+
n^{\rho}\epsilon_{\sigma \mu p n}=\epsilon_{\rho \sigma \mu n} \
\mbox{and}
\ n^{\mu}\epsilon_{\rho \sigma p s}-
n^{\sigma}\epsilon_{\rho \mu p s}+
n^{\rho}\epsilon_{\sigma \mu p s}=\epsilon_{\rho \sigma \mu s_T}
\end{eqnarray}
allow one to come to the standard gluonic operators,
used in the operator product expansion for spin-dependent
case \cite{AR}

\begin{eqnarray}
\label{momG}
\int_0^1 dx x^k \Delta G(x)=
{i^{k-1}\over{2M(sn)}}
\langle p,s|\tilde G_{\sigma \alpha}(0)
\partial ^{\mu_1}...\partial ^{\mu_{k-1}}
G_{\sigma \beta}(0) |p,s \rangle
n^{\alpha} n_{\beta} n_{\mu_1}...n_{\mu_{k-1}}~, \nonumber \\
\int_0^1 dx x^k  \Delta G_T(x)=
{i^{k-1}\over{2M(s^2)}}
\langle p,s|\tilde G_{\sigma \alpha}(0)
\partial ^{\mu_1}...\partial ^{\mu_{k-1}}
G_{\sigma \beta}(0) |p,s \rangle
s_T^{ \alpha} n_{\beta} n_{\mu_1}...n_{\mu_{k-1}}~,
\end{eqnarray}
where $\tilde G_{\sigma \alpha}={1 \over 2} \epsilon_{\sigma \alpha \mu \nu}
G_{\mu \nu}$.
Taking the totally symmetric part of the matrix element

\begin{eqnarray}
\label{matr}
i^{k-1}\langle p,s|\tilde G_{\sigma \alpha}(0)
\partial ^{\mu_1}...\partial ^{\mu_{k-1}}
G_{\sigma \beta}(0) |p,s \rangle =a_k S_{\alpha \beta \mu_1...\mu_{k-1}}
s^{\alpha}
 p^{\beta}   p^{\mu_1}... p^{\mu_{k-1}},
\end{eqnarray}
where $S$ denotes the total symmetrization and $a_k$ is the scalar constant,
one immediately obtains the relation

\begin{eqnarray}
\label{momWW}
\int_0^1 dx x^k \Delta G(x)=(k+1)\int_0^1 dx x^k  \Delta G_T(x)~,
\end{eqnarray}
which is equivalent to the WW formula:

\begin{eqnarray}
\label{xWW}
\Delta G_T(x)=\int_x^1 {\Delta G (z)\over z}dz~.
\end{eqnarray}

The existence of this relation is very natural because of the
similarity between quark  and gluon
density matrices (see eqs. (\ref{quark}) and (\ref{dm}) ).
Actually, such a relation is emerging due to the symmetry properties
of the operators and does not depend neither on the fields
(quark or gluons) nor on the coefficient function
(c.f. \cite{BK}).

Our present knowledge on $\Delta G(x)$, which is not very precise,
allows a great freedom, so several different parametrizations have been
proposed in the literature \cite{B1,B2,GS,GRSV}. In ref.[2], we have shown
in Figs.1a and 2a,
some possible gluon helicity distributions $x \Delta G(x)$ and in Figs.1b and
2b, the corresponding $x \Delta G_T(x)$ obtained by using (\ref{xWW}).
It is worth
recalling, from these pictures that, in all cases $x \Delta G(x)$ and
$x \Delta G_T(x)$ are rather similar in shape and magnitude.

Let us now move to the calculation of short-distance subprocess.
For this, it is instructive to compare the two terms in the
gluon density matrix (\ref{dm}).
While the longitudinal term is in fact a two-dimensional transverse
antisymmetric tensor and corresponds to the density matrix
of a circularly polarized gluon

\begin{eqnarray}
\label{londen}
\Delta G(x) \epsilon ^{\rho \sigma p n}=
\Delta G(x) \epsilon_{TT} ^{\rho \sigma}~ ,
\end{eqnarray}
the transverse polarization term generates the circular
polarization in the plane, defined by one transverse and one
longitudinal direction

\begin{eqnarray}
\label{trden}
M \Delta G_T(x) \epsilon ^{\rho \sigma s_T n}=
\Delta G_T(x) \epsilon_{TL} ^{\rho \sigma}~ ,
\end{eqnarray}
and therefore corresponds to the circular {\it transverse} polarization
of gluon. Such a polarization state is clearly impossible for
on-shell collinear gluons. They should have either nonzero virtuality,
or nonzero transverse momentum. Note that one of these effects is required to
have nonzero anomalous contribution to the first moment of the structure
function $g_1$\cite{EST,CCM}.
One may consider this similarity as supporting the
mentioned relations between $\Delta G_T$ and anomalous gluon contribution
\cite{ST95,rav}.

We should adopt the second possibility, namely a nonzero
transverse momentum, because the gluon remains on-shell
and the explicit gauge invariance is preserved. In this case,
the transverse polarization of nucleon may be converted to the longitudinal
circular polarization of gluon. The similar effect  was
discussed earlier for quarks \cite{Ratcl,kt}
and for photons in QED \cite{QED}.

To calculate now the asymmetry
in short-distance subprocess, it is enough to find
the effective longitudinal polarization by projecting the
transverse polarization onto the gluon momentum:

\begin{eqnarray}
\label{pro}
s_L={\vec s_T \vec k \over{|\vec k|}} =s_T {k_T \over k_L}~.
\end{eqnarray}

The partonic longitudinal-transverse and
double transverse asymmetries can be easily obtained from the
longitudinal one according to,

\begin{eqnarray}
\label{att}
\hat A_{LT}={ {k_{T1}}\over {k_{L1}}} \hat A_{LL},~ ~ \hat A_{TT}={ {k_{T1} k_{T2}}\over {k_{L1} k_{L2}}} \hat A_{LL}~.
\end{eqnarray}
By neglecting the transverse momentum dependence of $\hat A_{LL}$ one has

\begin{eqnarray}
\label{attq}
\hat A_{TL} \sim {\langle k_T \rangle \over {\sqrt{\hat s}}} \hat A_{LL},~ ~
\hat A_{TT} \sim {\langle k_T^2 \rangle \over {\hat s}} \hat A_{LL},~ ~
\end{eqnarray}
where $\hat s$ is the partonic c.m. energy.

The most promising effects, as it was mentioned above,
seem to be the diffractive
$J/\psi$ production, where the suppression factor should be
of the order of $0.3$. However, even the case of the longitudinal
polarization, the large value which seems to be provided by the
analogue of the Landau theorem, requires further investigation.

\section{Positivity bound for the condensate of longitudinal gluons}

\vspace{1mm}
\noindent

  Note that $\Delta G_T(x)$ may be also considered as an analogue
 of transversity for quark,
since for gluons there is no such a difference, caused for quarks by
the two Dirac projections (chiral-even and chiral-odd).
It is proportional to the following gluon-nucleon matrix element

\begin{eqnarray}
\label{hel}
\Delta G_T(x)=
\langle +,1|-,0\rangle
\equiv M_{+-}~ ,
\end{eqnarray}
where $(+,-)$ and $(1,0)$ are the nucleon and gluon helicities , respectively.

At the same time we have

\begin{eqnarray}
\label{hel1}
G_L(x)=
\langle +,0|+,0\rangle=
\langle -,0|-,0)\rangle
\equiv M_0 ~.
\end{eqnarray}

To establish the connection with the existing positivity relations,
let us consider
first the forward $\gamma^*p$ elastic scattering  ($\gamma^*$ is
a massive photon),which allows to calculate the DIS structure functions.
It is described in terms of
four helicity amplitudes: $M_0, M_{+-}$ defined above and
\begin{eqnarray}
\label{hel3}
M_+=\langle +,1|+,1\rangle, \, ~  M_-=\langle -,1|-,1\rangle ~ .
\end{eqnarray}

There is a well-known condition established long time ago by
Doncel and de Rafael \cite{DDR}, written in the form

\begin{eqnarray}
\label{DDR}
|A_2| \leq \sqrt{R}~,
\end{eqnarray}
where  $A_2$ is the usual transverse asymmetry and
$R=\sigma_L/\sigma_T$ is the standard ratio in DIS. It reflects a non-trivial
positivity condition one has on the photon-nucleon helicity amplitudes, which read using the above notations

\begin{eqnarray}
\label{pdis}
(M_{+-})^2 \leq 1/2(M_+ + M_-) M_0 ~.
\end{eqnarray}

One can apply this result to the similar case of
gluon-nucleon scattering, adding to the definitions (4,5,6),
 the following ones

\begin{eqnarray}
\label{helg}
G(x)=M_+ + M_- ~,~
\Delta G(x)=M_+ - M_- ~.
\end{eqnarray}

As a result, the positivity relation (8) leads to
\begin{eqnarray}
\label{ineq}
|\Delta G_T(x)| \leq \sqrt{1/2G(x)G_L(x)}~.
\end{eqnarray}

It is most instructive to use this relation to estimate
$G_L$ from below
.

\begin{eqnarray}
\label{bound}
G_L(x) \geq 2[\Delta G_T(x)]^2/G(x)  = 2\lambda(x)G(x)~ ,
\end{eqnarray}
where $\lambda(x)= [\Delta G_T(x)/G(x)]^2$.

Note that given the data on $R$ in DIS, one obtains from (7) an
{\it upper bound} on $|A_2|$, which is satisfied
by polarized DIS data \cite{AB} and is far from saturation.
However (\ref{bound}) provides a {\it lower bound} on $G_L(x)$
since $G(x)$ is known from unpolarized DIS  or direct photon
production and $\Delta G_T(x)$ can be evaluated \cite{ST97},
in the twist-two approximation if one  uses eq.(3).
One obtains $\lambda(x) \simeq 0.01$ for $x \simeq 0.1$,
or so and our lower bound gives $G_L(x) \geq 0.3$ or so.
For lower $x$ values, due to the rapid rise of $G(x)$, $\lambda(x)$ is
much smaller and, for example, for $x \simeq 10^{-3}$,
we find $G_L(x) \geq 10$ or so.
At the same time, for very large $x \to 1$,
as $\Delta G_T(x)$ is similar to $\Delta G(x)(1-x)$, $\lambda$ is
close to zero, and $G_L(x)/G(x) \sim 0$.

Such a relation is of special interest, since it relates, at least formally,  different twist structures. This is by no means surprising,
because, in the case of polarized DIS, if we would have known $A_2$ before
$R=\sigma_L/\sigma_T$,
we could have also estimated the latter from below.
Note that physically the existence of such a relation
is due to the fact, that transverse and longitudinal gluon distributions
are generated by the same source, the gluon mass.

\section{Conclusions}

\vspace{1mm}
\noindent

Up to now, we did not discuss the entire twist three effects,
emerging from the QCD long-range interactions. In the singlet channel,
they are described by the complicated
set of the three-gluon correlations \cite{Ji}.

They may lead to large single spin asymmetries in the fragmentation
region of the unpolarized particle \cite{Ji}. This effect was studied
quantitatively \cite {hamsha}, by making use of
a nonperturbative estimate of the correlations. The latter was
performed by considering a semi-classical long-wave gluonic field
\cite{sha2}.The contributions of these correlations in DIS were also studied
and the Burkhardt-Cottingham sum rule was found to be valid \cite{BET94}.

Another interesting manifestation of the three-gluon correlations
in spin physics \cite{ZH} was suggested recently. Namely,
it was shown that the nonperturbative
contribution of charmed quarks to the proton spin
is related to the matrix element of the operator
$G_{\mu \nu} \tilde G_{\nu \alpha} G_{\alpha \mu}$, which was estimated
to be
very large. The authors argue, that such strong effects are absent
in the vector channel, which is studied in the unpolarized scattering.

However, the positivity conditions may be applied here also,
and leads to the relation $|<p,s|\bar c(o) \gamma_\mu \gamma^5 c(z) |p,s>|
 \leq <p,s|\bar c(o) \gamma_\mu c(z) |p,s>$. It might not be a useful
 bound for the local matrix element, since the r.h.s. which corresponds
 to the first moment of the unpolarized sea distribution, may be infinite.
 However if
 we apply it to the $x$  distributions, one gets
 $|\Delta c(x)| \leq c(x)$.
 The quantity $c(x)$ (unpolarized intrinsic charm) is known to be small,
 and the most precise data comes from DIS at HERA. Therefore,
 the polarized charm should be small, in the very small $x$ region,
 which, by no means, has been investigated by the existing
 polarized DIS experiments and therefore has
 nothing to do with the present available data. If the analysis
\cite{ZH} turns out to be valid, this would lead to a dramatic increase of
 charmed quark contribution at small $x$.
 However, from the more recent evaluation of the nonperturbative
 polarized charm \cite{Shur},
 it results a much smaller contribution.

In the case where the correlations are not strong,
the kinematic approximation we used is rather natural.
It is also supported by the fact,
that all the current high-twist calculations are compatible with such
a "kinematical dominance" \cite{maul}.
In the singlet channel,
it results in the self-consistent picture of the massive gluons,
contributiong to the polarized and unpolarized structure functions.

We are indebted to P. S\"oding, J. Bl\"umlein and W.-D. Nowak for
warm hospitality at DESY-IfH Zeuthen. We are grateful to B.L.~Ioffe
and A.~Sch\"afer for valuable comments.

This investigation was supported in part by INTAS Grant 93-1180 and
by the Russian Foundation for Fundamental Investigations under
Grant 96-02-17631.

\end{document}